\documentclass{svmult}

\usepackage{makeidx}     
\usepackage{graphicx}    
\usepackage{multicol}    

\makeindex             

\begin{document}

\def\A{{\mathcal A}}
\def\B{{\mathbb B}}
\def\K{{\mathbb K}}
\def\L{{\mathcal L}}
\def\F{{\mathbb F}}
\def\N{{\mathbb N}}
\def\C{{\mathbb C}}
\def\R{{\mathbb R}}
\def\Z{{\mathbb Z}}
\def\Q{{\mathbb Q}}
\def\X{{\mathbb X}}
\def\al{\alpha}
\def\be{\beta}
\def\ga{\gamma}
\def\ep{\epsilon}
\def\S{{\Sigma}}
\def\Q{{\mathbb Q}}
\def\SG{{\goth S}}


\title*{Automata-Based Adaptive Behavior 
for Economic Modelling Using Game Theory}
\titlerunning{Automata-based adaptive behavior for economic modelling}
\author{
   Rawan Ghnemat\inst{1} \and Saleh Oqeili\inst{1} \and 
   Cyrille Bertelle\inst{2} \and 
   G\'erard~H.E.~Duchamp\inst{3}
}
\authorrunning{R. Ghnemat, S. Oqeili, 
   C. Bertelle and G.H.E. Duchamp}
\institute{
   Al-Balqa' Applied University, \\ 
   Al-Salt, 19117\\ 
   Jordan\\
   \texttt{rawan.ghnemat@gmail.com, saleh@bau.edu.jo} 
\and
   LITIS - University of Le Havre, \\ 
   25 rue Philippe Lebon \\
   76620 Le Havre cedex, France\\
   \texttt{cyrille.bertelle@gmail.com}
\and
   LIPN - University of Paris XIII, \\ 
   99 avenue Jean-Baptiste Cl\'ement \\ 
   93430 Villetaneuse, France\\
   \texttt{gheduchamp@gmail.com}
}

\maketitle

\begin{abstract}
In this paper, we deal with some specific domains of applications to game 
theory. This is one of the major class of models in the new approaches 
of modelling in the economic domain.
For that, we use genetic automata which allow to buid adaptive strategies for 
the players. We explain how the automata-based 
formalism proposed - matrix representation of automata with multiplicities - 
allows to define a semi-distance between the strategy behaviors.
 With that tools, we are able to generate an automatic processus to compute 
emergent systems of entities whose behaviors are represented by 
these genetic automata.
\end{abstract}


\vspace*{0.5in}
\vspace*{-\baselineskip}

\section{Introduction: Adaptive Behaviour Modeling for Ga\-me Theory}

Since the five last decades, game theory \index{game theory}
has become a major aspect in economic
sciences modelling \index{economic modelling} and in a great number of domains where 
strategical aspects has to be involved. 
Game theory is usually defined as a mathematical tool allowing to analyse 
strategical interactions between individuals. \\

Initially funded by mathematical researchers, J. von Neumann, E. Borel or E. 
Zermelo in 1920s, game theory increased in importance in the 
1940s with a major work by J. von Neumann and O. Morgenstern and then with the 
works of John Nash in the 1950s \cite{Eb}. 
John Nash has proposed an original equilibrium ruled by an adaptive criterium. 
In game theory, the Nash equilibrium is a kind of optimal strategy 
\index{strategy} for games 
involving two or more players, whereby the players reach an outcome 
to mutual advantage. 
If there is a set of strategies for a game with the property that no player can 
benefit by changing his strategy while the other players keep their 
strategies unchanged, then this set of strategies and the corresponding payoffs 
constitute a Nash equilibrium. \\

We can understand easily that the modelization of a player behavior needs some 
adaptive properties \index{adaptive properties}. 
The computable model corresponding to genetic automata are in this way a good 
tool to modelize such adaptive strategy \index{adaptive strategy}.\\


The plan of this paper is the following. In the next section, we present some 
efficient algebraic structures, the automata with multiplicities, which allow to 
implement powerful operators. We present in section 3, some topological 
considerations about the definition of distances between automata which induces a 
theorem of convergence on the automata behaviors.
Genetic operators are proposed for these automata in section 4. For that 
purpose, we show that the relevant ``calculus'' is done by matrix representions 
unravelling then the powerful capabilities of such algebraic structures.   
In section 5, we focus our attention on the "iterated prisonner dilemma" and we 
buid an original evolutive probabilistic automaton for strategy modeling, 
showing that genetic automata are well-adapted to model adaptive strategies. 
Section 6 shows how we can use the genetic automata developed previously to 
represent agent evolving in complex systems description.  An agent behavior 
semi-distance is then defined and allows to propose an automatic computation of 
emergent systems as a kind of self-organization detection.

\section{Automata from boolean to multiplicies theory (Automata with scalars)}


Automata\index{automata} are initially considered as theoretical tools. They are created in the 
1950's following the works of A. 
Turing who previously deals with the definition of an abstract "machine". 
The aim of the Turing machines \index{Turing machines}
is to define the boundaries for what a computing 
machine could do and what it could not do.\\ 

The first class of automata, called finite state automata 
\index{finite state automata} corresponds to simple 
kinds of machines \cite{Sc}. 
They are studied by a great number of researchers as abstract concepts for 
computable building.
In this aspect, we can recall the works of some linguist researchers, for 
example N. Chomsky who defined the study of formal grammars.\\ 

In many works, finite automata are associated to a recognizing operator which allows to 
describe a language \cite{BR,Ei}. 
In such works, the condition of a transition is simply a symbol taken from an 
alphabet. 
From a specific state $S$, the reading of a symbol $a$ allows to make the 
transitions which are labeled by $a$ and $\ come\ from S$ 
(in case of a deterministic automaton - a DFA - there is only one transition - 
see below). \index{deterministic Finite Automaton}
A whole automaton is, in this way, associated to a language, the recognized 
language, which is a set of words. 
These recognized words are composed of the sequences of letters of the alphabet 
which allows to go from a specific state called initial state, to another 
specific state, called final state.\\  

A first classification is based on the geometric aspect~: DFA (Deterministic 
Finite Automata) and NFA (Nondeterministic Finite Automata).
\index{deterministic finite automaton} \index{nondeterministic finite automaton}

\begin{itemize}
\item In Deterministic Finite Automata, for each state there is at most one 
transition for each possible input and only one initial state.
\item In Nondeterministic Finite Automata, there can be none or more than one 
transition from a given state for a given possible input.
\end{itemize} 

Besides the classical aspect of automata as machines allowing to recognize 
languages, another approach consists in associating to the automata a functional 
goal. 
In addition of accepted letter from an alphabet as the condition of a transition, we 
add for each transition an information which can be considered as an output data 
of the transition, the read letter is now called input data. 
We define in such a way an {\it automaton with outputs} or {\it weighted 
automaton}.\\
\index{automaton with outputs} \index{weighted automaton}

Such automata with outputs give a new classification of machines. 
{\it Transducers} are such a kind of machines, they generate outputs based on a 
given input and/or a state using actions.
\index{transducers} 
They are currently used for control applications. 
{\it Moore machines} are also such machines where output depends only on a 
state, i.e. the automaton uses only entry actions. 
\index{Moore machines}
The advantage of the Moore model is a simplification of the behaviour.\\ 

Finally, we focus our attention on a special kind of automata with outputs which 
are efficient in an operational way. 
This automata with output are called {\it automata with multiplicities}.
\index{automaton with mutiplicities} 
An automaton with multiplicities is based on the fact that the output data of 
the automata with output belong to a specific algebraic structure, a 
semiring \cite{Go,St}. 
In that way, we will be able to build effective operations on such automata, 
using the power of the algebraic structures of the output data 
and we are also able to describe this automaton by means of a matrix 
representation with all the power of the new (i.e. with semirings) linear algebra.\\

\begin{definition}
{\bf (Automaton with multiplicities)}\\
 An automaton with multiplicities over an alphabet  $A$  and a semiring $K$ is 
the 5-uple $(A,Q,I,T,F)$ where
\begin{itemize}
\item  $Q=\{S_1,S_2\cdots S_n\}$ is the finite set of state;
\item   $I: Q\mapsto K$  is a function over the set of states, which 
associates to each initial state a value of K, called entry cost, and to non-
initial state a zero value ;
\item	$F: Q\mapsto K$  is a function over the set states, which 
associates to each final state a value of K, called final cost, and to non-final 
state a zero value;
\item $T$ is the transition function, that is $T: Q\times A\times Q\mapsto K$ 
which to a state $S_i$, a letter $a$ and a state $S_j$ associates a value $z$ of 
$K$ (the cost of the transition) 
if it exist a transition labelled with $a$ from the state $S_i$  to the 
state $S_j$ and and zero otherwise.\\   
\end{itemize}
\end{definition}

\begin{remark} 
Automata with multiplicities are a generalisation of finite automata. In fact, 
finite automata can be considered as automata with multiplicities in the 
semiring $K$, the boolan set $B=\{0,1\}$ (endowed with the logical ``or/and'').
 To each transition we affect 1 if it exists and 0 if not.\\
\end{remark}

\begin{remark}
We have not yet, on purpose, defined what a semiring is. Roughly it is the least 
structure 
 which allows the matrix ``calculus'' with unit (one can think of a ring without the 
"minus" operation).
The previous automata with multiplicities can be, equivalently, expressed by a 
matrix representation which is a triplet 
\begin{itemize}
\item $\lambda\in K^{1\times Q}$ which is a row-vector which coefficients are 
$\lambda_i=I(S_i)$,
\item $\gamma\in K^{Q\times 1}$ is a column-vector which coefficients are 
$\gamma_i=F(S_i)$,
\item $\mu: A^*\mapsto K^{Q\times Q}$ is a morphism of monoids (indeed 
$K^{Q\times Q}$ is endowed with the product of matrices) such that 
the coefficient on the $q_i$th row and $q_j$th column of $\mu(a)$ is 
$T(q_i,a,q_j)$ 
\end{itemize} 

\end{remark}

\section{Topological considerations}

If $K$ is a field, one sees that the space $\A_{(n)}$ of automata of dimension 
$n$ (with multiplicities in $K$) is a $K$-vector space of dimension 
$k.n^2+2n$ ($k$ is here the number of letters). So, in case the ground field is 
the field of real or complex numbers \cite{Bo1}, one 
can take any vector norm (usually one takes one of the H\"older norms 
$||(x_i)_{i\in I}||_\alpha := \big(\sum_{i\in I} | x_i 
|^\alpha\big)^{\frac{1}{\alpha}}$ 
for $\alpha\geq 1$,
 but any norm will do) and the distance is derived, in the classical way, by 
\begin{equation}
d(\A_1,\A_2)=norm(V(\A_1)- V(\A_2))
\end{equation}
where $V(\A)$ stands for the vector of all coefficients of $\A=(\lambda,\mu,\gamma)$ 
arranged in some order one has then the result of Theorem \ref{th1}.
 Assuming that $K$ is the field of 
real or complex numbers, we endow the 
space of series/behaviours with the topology of  pointwise convergence (Topology 
of F. Treves \cite{Tr}).  

\begin{theorem}\label{th1}
Let  $(\A_n)$ be a sequence of automata with limit $\L$ ($\L$ is an automaton), 
then one has
\begin{equation}
Behaviour(\L)=\lim_{n\rightarrow \infty} Behaviour(\A_n)
\end{equation}
where the limit is computed in the topology of Treves.
\end{theorem}

\section{Genetic automata as efficient operators}

We define the chromosome for each automata with multiplicities as the sequence 
of all the matrices 
 associated to each letter from the (linearly ordered) alphabet. 
The chromosomes are composed with alleles 
which are here the lines of the matrix \cite{BFJOP2}.\\

In the following, genetic algorithms are going to generate new automata 
containing possibly new transitions 
from the ones included in the initial automata.\\

The genetic algorithm over the population of automata with multiplicities 
follows a reproduction iteration 
broken up in three steps \cite{Gol,Mi,Ko}:
\begin{itemize}
\item 	{\it Duplication}: where each automaton generates a clone of itself; 
\item	{\it Crossing-over}: concerns a couple of automata. Over this couple, we 
consider a sequence of lines of each 
matrix  for all. For each of these matrices, a permutation on the lines of the 
chosen 
sequence is made between the analogue matrices of this couple of automata; 
\item	{\it Mutation}: where a line of each matrix  is randomly chosen and a sequence 
of new values is given for 
this line.
\end{itemize}

Finally the whole genetic algorithm scheduling for a full process of 
reproduction over all the population of 
automata is the evolutionary algorithm:

\begin{enumerate}
\item For all couple of automata, two children are created by duplication, 
crossover and mutation 
mechanisms;
\item The fitness for each automaton is computed;
\item For all 4-uple composed of parents and children, the performless automata, 
in term of fitness 
computed in previous step, are suppressed. The two automata, still living, 
result from the 
evolution of the two initial parents.
\end{enumerate}

\begin{remark}
The fitness is not defined at this level of abstract formulation, but it is 
defined 
corresponding to the context for which the automaton is a model, as we will do in 
the next section.
\end{remark}

\section{Applications to competition-cooperation modeling using prisoner 
dilemma}

We develop in this section how we can modelize competition-cooperation processes 
in a same automata-based representation. The genetic computation allows to make 
automatic transition from competition to cooperation or from coopeartion to 
competition. The basic problem used for this purpose is the well-known prisoner 
dilemma \cite{Ax}.

\subsection{From adaptive strategies to probabilistic automata}

The prisoner dilemma is a two-players game where each player has two possible 
actions: 
cooperate ($C$) with its adversary or betray him ($\overline{C}$). So, four 
outputs are possible for the global 
actions of the two players. A relative payoff is defined relatively to these 
possible outputs, as 
described in the following table where the rows correspond to one player 
behaviour and the 
columns to the other player one.\\

\begin{table}[htp]
        \begin{center}
        \begin{tabular}{|l|c|c|} \hline
                                & $C$           & $\overline{C}$ \\ \hline
                $C$             & (3,3)     & (0,5)       \\ \hline
                $\overline{C}$  & (5,0)     & (1,1)       \\ \hline
        \end{tabular}
        \caption{Prisoner dilemma payoff}
        \label{prisonerDilemmaPayoff}
\end{center}
\end{table}

In the iterative version of the prisoner's dilemma, successive steps can be defined.  
Each player do not know the action of its adversary during the current step but 
he knows it for the preceding step. 
So, different strategies can be defined for a player behaviour, the goal of each 
one is to 
obtain maximal payoff for himself.\\
 
\begin{figure} [htp]
\begin{center}
\includegraphics[scale=0.7]{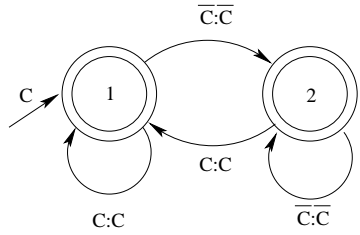}
\caption{Tit-for-tat strategy automaton}
\label{titfortat}
\end{center}
\end{figure}

\begin{figure} [htp]
\begin{center}
\includegraphics[scale=0.7]{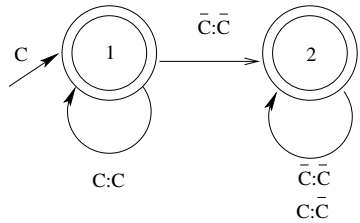}
\caption{Vindictive strategy automaton}
\label{vindictive}
\end{center}
\end{figure}

In Figures \ref{titfortat} and \ref{vindictive}, we describe two strategies 
with transducers. 
Each transition is labeled by the input corresponding to the player perception 
which is the precedent adversary action and the output corresponding to the 
present player 
action.  
The only inital state is the state 1, recognizable by the incoming arrow labeled 
only by the output. 
The final states are the states 1 and 2, recognizable with the double circles.\\
 
\begin{figure} [htp]
\begin{center}
\includegraphics[scale=0.7]{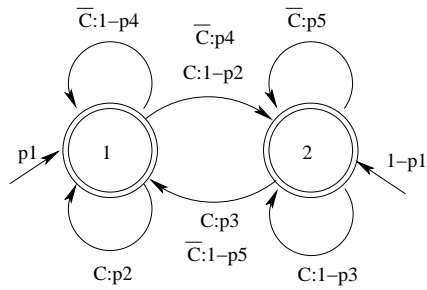}
\caption{Probabilistic multi-strategies two-states automaton}
\label{probaDilemma}
\end{center}
\end{figure}

In the strategy of Figure \ref{titfortat}, the player has systematically the 
same behaviour as its adversary at the previous step. 
In the strategy of Figure \ref{vindictive}, the player chooses definitively to 
betray as soon as his adversary does it.  
The previous automaton represents static strategies and so they are not well 
adapted for the 
modelization of evolutive strategies. 
For this purpose, we propose a model based on a probabilistic automaton 
described by Figure \ref{probaDilemma} \cite{BFJOP1}.\\

This automaton represents all the two-states strategies for cooperation  and 
competitive   
behaviour of one agent against another in prisoner's dilemma.\\

The transitions are labeled in output by the probabilities $p_i$ of their 
realization.  The first state is the state reached after cooperation action and 
the second state is reached after betrayal. \\

For this automaton, the associated matrix representation, as described 
previously, is:

\begin{eqnarray} 
I &=& \left( \begin{array}{cc} p_1 & 1-p_1 \end{array} \right); \\
F &=& \left( \begin{array}{c} p_6 \\ 1-p_6 \end{array} \right);\\
T(C) &=& \left( \begin{array}{cc} p_2 & 1-p_2\cr p_3 & 1- p_3 \end{array} \right);\\ 
T(\overline{C}) &=& \left( \begin{array}{cc} 1-p_4 & p_4\cr 1-p_5 & p_5 \end{array} \right)
\end{eqnarray}

\subsection{From probabilistic automata to genetic automata}

With the matrix representation of the automata, we can compute genetic automata 
as described 
in previous sections. Here the chromosomes are the sequences of all the matrices 
associated to 
each letter. We have to define the fitness in the context of the use of these 
automata. The 
fitness here is the value of the payoff.

\subsection{General Genetic Algorithm Process for Genetic Automata}

A population of automata is initially generated. These automata are playing 
against a predefined 
strategy, named $S_0$.\\

Each automaton makes a set of plays. At each play, we run the probabilistic 
automaton 
which gives one of the two outputs: ($C$) or ($\overline{C}$). With this output 
and the $S_0$'s output, we compute the payoff of the automaton, according with 
the payoff table.\\

At the end of the set of plays, the automaton payoff is the sum of all the 
payoffs of each play. 
This sum is the fitness of the automaton. At the end of this set of plays, each 
automaton has 
its own fitness and so the selection process can select the best automata. At 
the end of these 
selection process, we obtain a new generation of automata.\\

This new generation of automata is the basis of a new computation of the 3 
genetics operators.\\

This processus allows to make evolve the player's behavior which is modelized by 
the probabilistic multi-stra\-te\-gies two-states automaton from cooperation to 
competition or from competition to cooperation. The evolution of the strategy is 
the expression of an adaptive computation. This leads us to use this formalism 
to implement some self-organisation processes which occurs in complex systems.  


\section{Extension to Emergent Systems Modeling}

In this section, we study how evolutive automata-based modeling can be used to 
compute automatic emergent systems. The emergent systems have to be understood 
in the meaning of complex system paradigm that we recall in the next section. We 
have previously defined some way to compute the distance between automata and we 
use these principles to define distance between agents behaviours that are modeled 
with automata. Finally, we defined a specific fitness that allows to use genetic 
algorithms as a kind of reinforcement method which leads to emergent system 
computation \cite{Ho}.

\subsection{Complex System Description Using Automata-Ba\-sed Agent Model}

\begin{figure*} [ht]
\begin{center}
\includegraphics[scale=0.85]{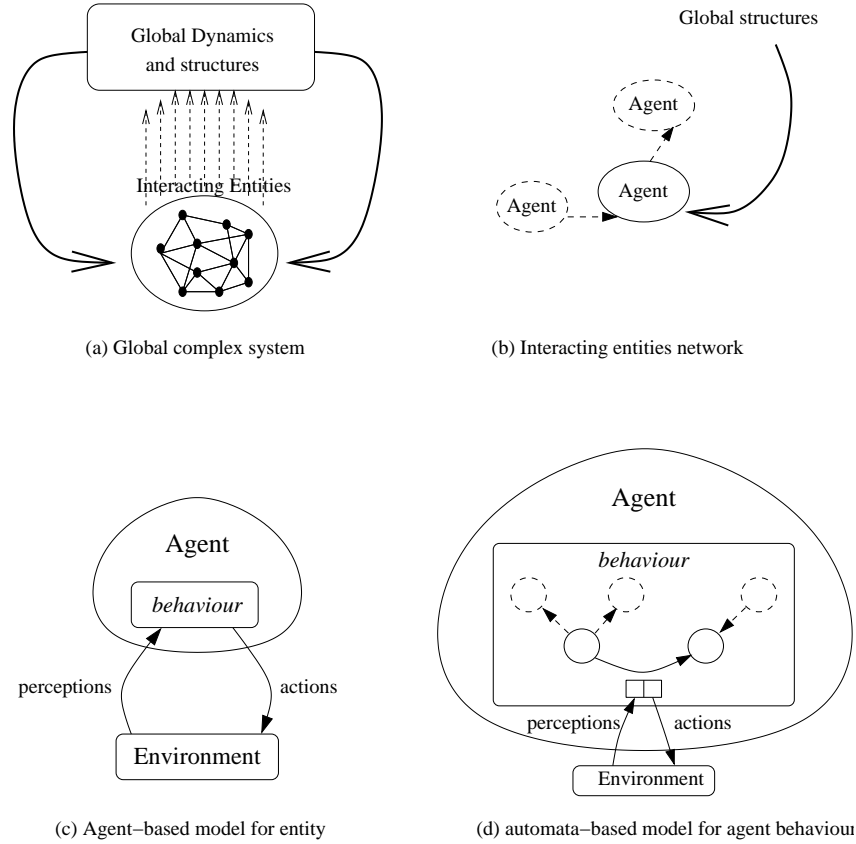}
\caption{Multi-scale complex system description: from global to individual 
models}
\label{sys2beh}
\end{center}
\end{figure*}

According to General System Theory \cite{gst, Mo}, a complex system is composed 
of entities in mutual interaction and interacting with the outside environment. A 
system has some characteristic properties which confer its structural aspects, 
as schematically described in part (a) of Figure \ref{sys2beh}: 
\begin{itemize}
\item The set elements or entities are in interactive dependance. The alteration 
of only one entity or one interaction reverberates on the whole system. 
\item A global organization emerges from interacting constitutive elements. This 
organization can be identified and carries its own autonomous behavior while it 
is in relation and dependance with its environment. The emergent organization 
possesses new properties that its own constitutive entities don't have. "The 
whole is more than the sum of its parts". 
\item The global organization retro-acts over its constitutive components. "The 
whole is less than the sum of its parts" after E. Morin.\\
\end{itemize}

The interacting entities network as described in part (b) of Figure 
\ref{sys2beh} leads each entity to perceive informations or actions from other 
entities or from the whole system and to act itself.\\

A well-adapted modeling consists of using an agent-based representation which is 
composed of the entity called agent as an entity which perceives and acts on an 
environment, using an autonomous behaviour as described in part (c) of 
Figure \ref{sys2beh}.\\

To compute a simulation composed of such entities, we need to describe the 
behaviour of each agent. This one can be schematically described using internal 
states and transition processes between these states, as described in part 
(d) of Figure \ref{sys2beh}.\\  

There are several definitions of ``agents'' or ``intelligent agents'' according 
to their behaviour specificities~\cite{Fe, We}.
Their autonomy  means that the agents try to satisfy a goal
and execute actions, optimizing a satisfaction function to reach it.\\

For agents with high level autonomy, specific actions are realized even when no 
perception are detected from the environment. 
To represent the process of this deliberation, different formalisms can be 
used and a behaviour decomposed in internal states is an effective approach.
Finally, when many agents operate, the social aspects must also be taken into 
account. 
These aspects are expressed as communications through agent organisation with 
message
passing processes. 
Sending a message is an agent action and receiving a message is an agent 
perception. The previous description based on the couple: perception and action, 
is well adapted to this.

\subsection{Agent Behavior Semi-Distance}

We describe in this section the bases of the genetic algorithm used on the 
probabilistic automata allowing to manage emergent self-organizations in the 
multi-agent simulation.\\

For each agent, we define $e$ an evaluation function of its own
behaviour returning the matrix $M$ of values such that $M_{i,j}$ is 
the output series from all possible successive perceptions when
starting from the initial state $i$ and ending at the final state $j$,
without cycle. It will clearly be $0$ if either $i$ is not an initial
state or $j$ is not a final one and the matrix $M_{i,j}$ is indeed a matrix of 
evaluations \cite{BR} of subseries of 
\begin{equation}
M^*:=(\sum_{a\in A} \mu(a)a)^*
\end{equation}

Notice that the
coefficients of this matrix, as defined, are computed whatever the
value of the perception in the alphabet $A$ on each transition on the successful
path\footnote{A {\it succesful path} is a path from an initial state to a final 
state}. That means that the contribution of the agent behaviour for
collective organization formation is only based, here, on
probabilities to reach a final state from an initial one.     
This allows to preserve individual characteristics in each agent
behaviour even if the agent belongs to an organization.\\

Let $x$ and $y$ two agents and $e(x)$ and $e(y)$ their respective
evaluations as described above.
We define $d(x,y)$ a semi-distance (or pseudometrics, see \cite{Bo1} ch IX) 
between the two agents $x$ and $y$ as 
$||e(x)-e(y)||$, a matrix norm of the difference of their
evaluations. Let ${\cal{V}}_x$ a
neighbourhood of the agent $x$, relatively to a specific criterium, for
example a spatial distance or linkage network.  
We define $f(x)$ the agent fitness of the agent $x$ as~:
$$
f(x) = 
\left\lbrace
\begin{array}{ll}
\frac{ {\displaystyle card({\cal{V}}_x) } }
     { {\displaystyle \sum\limits_{y_i \in {\cal{V}}_{x}} d(x, y_i)^2} } 
     \ \ \ \ &\mbox{if} \sum\limits_{y_i \in {\cal{V}}_{x}} d(x,
y_i)^2 \neq 0 \\
\infty &\mbox{otherwise}
\end{array}
\right.
$$

\subsection{Evolutive Automata for Automatic Emergence of Self-Organized Agent-
Based Systems}
 
In the previous computation, we defined a semi-distance between two agents. This 
semi-distance is computed using the matrix representation 
of the automaton with multiplicities associated to the agent behaviour. This 
semi-distance is based on successful paths computation which 
needs to define initial and final states on the behaviour automata. For specific 
purposes, we can choose to define in some specific way, the 
initial and final states. This means that we try to compute some specific action 
sequences which are chararacterized by the way of going from some specific 
states (defined here as initial ones) to some specific states (defined here as 
final ones).\\

Based on this specific purpose which leads to define some initial and final 
states, we compute a behaviour semi-distance and then the fitness function defined 
previously. This fitness function is an indicator which returns high value when 
the evaluated agent is near, in the sense of the behaviour semi-distance defined 
previously, to all the other agents belonging to a predefined neighbouring.\\

Genetic algorithms will compute in such a way to make evolve an agent population 
in a selective process. So during the computation, the genetic algorithm will make 
evolve  the population towards a newer one with agents more and more adapted to the 
fitness. The new population will contain agents with better fitness, so the 
agents of a population will become nearer each others in order to improve their fitness. 
In that way, the genetic algorithm reinforces the creation of a system which 
aggregates agents with similar behaviors, in the specific way of the definition 
of initial and final states defined on the automata.\\

The genetic algorithm proposed here can be considered as a modelization of the 
feed-back of emergent systems which leads to gather agents of similar behaviour, 
but these formations are dynamical and we cannot predict what will be the 
set of these aggregations which depends of the reaction of agents during the 
simulation. Moreover the genetic process has the effect of generating a feed-
back of the emergent systems on their own contitutive elements in the way that 
the fitness improvement lead to bring closer the agents which are picked up 
inside the emergent aggregations.\\

For specific problem solving, we can consider that the previous fitness 
function can be composed with another specific one which is able to measure the 
capability of the agent to solve one problem. This composition of fitness 
functions leads to create emergent systems only for the ones of interest, that 
is, these systems are able to be developed only if the aggregated agents are 
able to satisfy some problem solving evaluation. 

\section{Conclusion}

The aim of this study is to develop a powerful algebraic structure to represent 
behaviors concerning cooperation-competition processes and on which we can 
add genetic operators. We have explained how we can use these structures for 
modeling adaptive behaviors needed in game theory. More than for this 
application, we have described how we can use such adaptive computations to 
automatically detect emergent systems inside interacting networks of entities 
represented by agents in a simulation.


\end{document}